\documentstyle[twocolumn,aps,prd]{revtex}
\tighten
\begin{document}
\draft
%\usepackage{amstex}   % useful for coding complex math
%\mathindent\parindent % needed in case "Amstex" is used

\title{The Gravitational Radiation degrees of freedom in Hyperbolic Systems
for Numerical Relativity}
\author{C.~Bona, C.~Palenzuela}
\address{Departament de Fisica. Universitat de les Illes Balears.
     Ctra de Valldemossa, km 7.5. 07071 Palma de Mallorca. Spain.}

\maketitle

\begin{abstract}
The gravitational radiation degrees of freedom of freedom are 
described in the framework of the 3+1 decomposition of spacetime.
The relationship with eigenfields of the Kidder-Scheel-Teukolsky
(KST) equations \cite{Cornell01} is established. This relationship
is used to fix a parameter in the KST equations which is related
to the ordering ambiguity of space derivatives in the Ricci tensor,
which is inherent to first order evolution systems, like the ones
currently used in Numerical Relativity applications.
\end{abstract}

\section{Introduction}

The structure of Einstein field equations has deserved great
interest since the very beginning of General Relativity. It was
early noticed that, by rearranging the order or partial
derivatives, the principal part of the four dimensional Ricci
tensor could be written as a sort of generalized wave equation
\cite{DeDonder,Lanczos}
\begin{equation}\label{Ricci4D}
   2\;R_{\mu\nu}=-\Box g_{\mu\nu} + \partial_{\mu}\Gamma_{\nu}
                 + \partial_{\nu}\Gamma_{\mu}+...
\end{equation}
where the box stands for the d'Alembert operator acting on
functions and we have written for short
\begin{equation}\label{Gamma4D}
   \Gamma^{\mu}\equiv g^{\rho\sigma}\;{\Gamma^{\mu}}_{\rho\sigma}
    = -\Box x^{\mu}\;\;.
\end{equation}

This opened the way to the use of harmonic spacetime coordinates
($\Box x^{\mu}=0$) in order to obtain an hyperbolic evolution
system \cite{Choquet55,Hawking,DeTurck}.

By the middle of the past century, however, the interest focused
in the relativistic Cauchy problem. The 3+1 decomposition of the
line element:
\begin{eqnarray}\label{metric4D}
    ds^2 &=& - \alpha^2\;dt^2  \nonumber \\
    &+& \gamma_{ij}\;(dx^i+\beta^i\;dt)\;(dx^j+\beta^j\;dt)
     \;\;\;\;i,j=1,2,3
\end{eqnarray}
allowed one to express six of the ten second order original
equations as a system of evolution equations for the metric
$\gamma_{ij}$ and the extrinsic curvature $K_{ij}$ of the
$t=constant$ slices, namely:
\begin{eqnarray}
   (\partial_t -{\cal L}_{\beta})\gamma_{ij} &=& -2\;\alpha\;K_{ij}
\label{evolve_metric} \\
   (\partial_t -{\cal L}_{\beta}) K_{ij} &=& -\nabla_i\alpha_j
       + \alpha\; [{}^{(3)}R_{ij}-2K^2_{ij}+ trK\;K_{ij}]
\label{evolve_curv}
\end{eqnarray}
where $\cal L $ stands for the Lie derivative (we restrict
ourselves to the vacuum case for simplicity). The remaining four
equations could instead be expressed as constraints:
\begin{eqnarray}
   ^{(3)}R - tr(K^2) + (trK)^2 &=& 0
\label{energy_constraint}  \\
   \nabla_k\:{K^k}_{i}-\partial_i(trK) &=& 0 \;\;\;.
\label{momentum_constraint}
\end{eqnarray}

This opened the door to a new way of obtaining hyperbolic
evolution systems \cite{Choquet83,Bona92,Frittelli94}. The key
point was to use in one or another way the momentum constraint
(\ref{momentum_constraint}) to ensure hyperbolicity while keeping
the freedom of choosing arbitrary space coordinates on every
$t=constant$ slice. This allows for instance to use normal
coordinates ($\beta^{i}=0$) without affecting the mathematical
structure of the evolution system.(See Ref. \cite{Bona99} for a
detailed comparison of the "old" and "new" hyperbolic systems).

These findings came at the right moment for people working on
Numerical Relativity with a view on the gravitational waves
detector projects starting by the turn of the century. Following
the wake of these first works, many groups found their own way of
combining the momentum constraint with the evolution equation
(\ref{evolve_curv}), leading in each case to a new brand of
hyperbolic systems
\cite{Cornell01,Choquet95,Bona95,Bona97,Frittelli96,Friedrich96,Putten96,Estabrook97,Iriondo97,Bonilla98,Stewart98,Yoneda99,Yoneda00,Yoneda01,Alcubierre99,Anderson99}.

Suddenly, the problem of having too few hyperbolic formalisms for
Numerical Relativity turned into the opposite problem of having
too many of them. Some of the works considered even multiparameter
families of hyperbolic systems
\cite{Cornell01,Bona99,Frittelli96}. Faced with the problem of
choice, we think that the right question at this point is: what
are hyperbolic systems for?

There are many answers, of course, but we can get a  hint when we 
realise that hyperbolic systems basically describe propagation 
phenomena along characteristic lines. This sets up the question of
whether we can use these formalisms in the context of gravitational
wave propagation. The 3+1 formalism does not seem at first sight
very useful in that context, because it describes spacetime as a 
foliation of $t=constant$ hypersurfaces. Then, the natural structure
associated with gravitational waves in that formalism is not the
light cone itself, but the two-dimensional surfaces (wavefronts)
obtained as the intersection of the light cone with the $t=constant$
hypersurfaces. The geometry of wavefronts will be discussed in Section
II. The second difficulty arises from the fact that, because of the
constraint equations (\ref{energy_constraint}, \ref{momentum_constraint}),
only two degrees of freedom correspond to gravitational radiation in
spite of the complete (strong) hyperbolicity of the formalisms.
This point can be easily illustrated by looking at the list of 
characteristic speeds given in the KST paper \cite{Cornell01} namely
\{$0, \pm 1, \pm c_1, \pm c_2, \pm c_3$\} where $c_1$, $c_2$ and $c_3$
depend of a list of arbitrary parameters related with the gauge choice
($c_1$) and with the way in which the constraints are used to enforce
hyperbolicity.
It is clear that the degrees of freedom propagating with speed $0$,
$\pm c_1$, $\pm c_2$ or $\pm c_3$ can not describe gravitational
radiation, which we know that propagates at light speed ($v=\pm 1$),
independently of how we manage to write down our equations. The two
remaining degrees of freedom, as expected, are related with transverse
traceless components of the extrinsic curvature $K_{ij}$. We will analyse
more closely these eigenfields in Section III.

\section{Light cones and wavefronts}

Let us consider the propagation of a burst of gravitational waves. From the
geometrical point of view, it can be described as a foliation of spacetime
by a set of null hypersurfaces (light cones), namely
\begin{eqnarray}
   \phi (x,t) &=& constant 
\label{wavefront1}\\ 
   d\phi \cdot d\phi &=& 0
\label{perpcondition}   
\end{eqnarray}
From every $t=constant$ hypersurface, the propagation is seen instead as a
succession of wavefronts. It can be obtained by setting $t=t_o$ in 
(\ref{wavefront1}) to obtain  
\begin{equation}\label{wavefront2}
   \phi (x,t_o) = constant
\end{equation}
so that the null foliation (\ref{wavefront1}) induces a spacelike foliation 
(\ref{wavefront2}) by two-dimensional surfaces (wavefronts) on every
$t=constant$ slice.

Let us consider now an adapted coordinate system on the $t=t_o$ slice so
that the space unit normal $n_k$ to the wavefront (\ref{wavefront2}) is
given by
\begin{equation}\label{normal}
    n_k = N\;{\delta_k}^z 
\end{equation}
and the other two coordinates $x^a$ ($a=1,2$) display the wavefronts surface.
The line element for the $t=t_o$ slice can then be written as (2+1 decomposition)
\begin{eqnarray}\label{metric3D}
    \gamma_{ij}\;&dx^i&\;dx^j = N^2\;dz^2 \nonumber \\
    &+& \sigma_{ab}\;(dx^a+\lambda^a\;dz)\;(dx^b+\lambda^b\;dz) \;\;\;a,b=1,2
\end{eqnarray}
where $\sigma_{ab}$ is the induced metric on every wavefront
surface, so that it describes the intrinsic geometry of the wavefronts. 

The extrinsic geometry of the wavefront foliation (\ref{wavefront2}) is in
turn described by the second fundamental form $\kappa_{ab}$, namely
\begin{equation}\label{curv2D}
    \kappa_{ab} = 
     \frac{1}{2\;N}\;(\partial_z-{\cal L}_{\lambda})\sigma_{ab}\:.
\end{equation}
Both $\sigma_{ab}$ and $\kappa_{ab}$ are 2+1 tensors, in the sense that they
transform in a covariant way under changes of coordinates of the form
\begin{eqnarray}\label{transformation}
   \tilde{z}&=&f(z) \\
   {\tilde{x}}^a&=&h^a(x,z)
\end{eqnarray}
which transform the wavefront foliation (\ref{wavefront2}) into itself.

Notice that $\sigma_{ab}$, $\kappa_{ab}$ can be easily expressed in terms
of the three-dimensional metric $\gamma_{ij}$ and the spacelike unit
normal $n_k$, namely
\begin{eqnarray}
   \sigma_{ab} &=& \gamma_{ab}
\label{correspondence1} \\
   \kappa_{ab} &=& - n_k \; {\Gamma^k}_{ab}
\label{correspondence2} 
\end{eqnarray}
where ${\Gamma^k}_{ij}$ stand for the Christoffel symbols of $\gamma_{ij}$.

\section{Solving the ordering ambiguity}

Let us come back now to the results of the KST paper \cite{Cornell01}. 
The principal part of the first order evolution systems presented there 
can be written in Flux-Conservative form for the array of variables 
$u=\{\gamma_{ij},K_{ij},d_{kij}\}$, namely
\begin{equation}\label{fluxconservative}
  \frac{1}{\alpha} \partial_t u + \partial_k F^k(u) =...
\end{equation}
so that propagation along the direction given by $n_k$ can be studied by
solving the following characteristic eigenvalue problem
\begin{equation}\label{RH}
   F^n(u) \equiv n_k\:F^k(u) = v\:u
\end{equation}
From the transverse components $K_{ab}$, one has \cite{Cornell01}
\begin{eqnarray}\label{fluxK}
    F^n(K_{ab}) &=& \frac{1}{2} \;[{d^n}_{ab} - (1+\zeta)\;{d_{(ab)}}^n \nonumber \\
    &+& m\; \gamma_{ab}\; ({d^{nk}}_k - {d_k}^{kn})]
\end{eqnarray}
so that the parameter $m$ (corresponding to $\gamma$ in \cite{Cornell01})
does not contribute to the transverse traceless part. On the other hand, 
one has \cite{Cornell01}
\begin{eqnarray}
    F^n(d_{nab}) &=& 2\;K_{ab} - \chi\; \gamma_{ab}\; (K^{nn} - trK)
\label{fluxesd1} \\
    F^n(d_{abn}) &=& - \frac{\eta}{2}\gamma_{ab}\; (K^{nn} - trK)
\label{fluxesd2}    
\end{eqnarray}
so that it is clear again that the parameters $\eta$, $\chi$ do not contribute
to the transverse traceless part.

This means that the traceless part of the combinations
\begin{equation}\label{ligtheigenfields}
  K_{ab} \pm \frac{1}{2}\;[{d^n}_{ab} - (1+\zeta)\; {d_{(ab)}}^n]
\end{equation}
corresponds to eigenfields propagating with light speed ($v=\pm 1$, respectively).
The parameter $\zeta$ appearing in (\ref{ligtheigenfields}) is related with the
ordering ambiguity inherent to first order formalisms, where the space derivatives
$d_{kij}$ are considered to be independent quantities, so that the identity
\begin{equation}\label{derivative_constraint}
  \partial_{[r} d_{s]ij} = 0
\end{equation}
is not granted and must be considered as a supplementary constraint
\cite{Cornell01}. The value $\zeta=-1$ corresponds to a decomposition
of the three-dimensional Ricci tensor of the form (\ref{Ricci4D}), which is
closer to the wave equation. The value $\zeta = +1$ corresponds instead to 
the simpler decomposition
\begin{equation}\label{Ricci3D}
   {^{(3)}R}_{ij}= \partial_k{\Gamma^k}_{ij} -\partial_{(i}{\Gamma^k}_{j)k}
                 + {\Gamma^k}_{kr}\;{\Gamma^r}_{ij}
                 - {\Gamma^k}_{ri}\;{\Gamma^r}_{kj}
\end{equation}
Notice that, allowing for (\ref{correspondence2}), the combinations
(\ref{ligtheigenfields}) in the $\zeta = +1$ case can be written as
\begin{equation}\label{tensoreigenfields}
  K_{ab} \pm \kappa_{ab}
\end{equation}
so that they have a geometrical meaning in terms of the extrinsic curvature
of the $t=constant$ hypersurfaces and the wavefront foliation. For any other
value of $\zeta$, however, the eigenfields described by the traceless
part of (\ref{ligtheigenfields}) are not covariant under the coordinate
transformations (\ref{transformation}) and do not admit then any
consistent, coordinate independent, physical interpretation.
It follows that one can solve the ordering ambiguity for the 
derivatives in the principal part of the Ricci tensor by requiring
that the eigenfields corresponding to the transverse traceless degrees
of freedom do have a clear geometrical meaning in terms of the geometry
of the $t=constant$ hypersurfaces and that of the wavefronts. 
This requirement amounts to select the particular value $\zeta=+1$,
corresponding to the classical decomposition (\ref{Ricci3D}).

\section{Concluding remarks}

The parameter dependence of the eigenfields (\ref{ligtheigenfields})
may seem surprising at the first sight, because these transverse
traceless degrees of freedom are related with a real physical
phenomenon (gravitational waves), which is independent on how we do
actually decompose the Ricci tensor. This paradox can be solved by
noticing that the Fluxes appear in the evolution equations
(\ref{fluxconservative}) under a divergence operator; the parameter
$\zeta$ appears when including the "rotational" constraint
(\ref{derivative_constraint}), but all these terms just cancel under
the divergence operator, so that all values of $\zeta$ are physically
equivalent.

The special choice $\zeta = +1$ corresponds to the only case in which
the Fluxes themselves inherit the covariance properties of the equations.
We claim that this is an important feature, because then gravitational
radiation features are explicitly inherited by the eigenfields of
the system.

On the other hand, there are indications that $\zeta = +1$ may not be
convenient from the Numerical Relativity point of view. Most of the 
formalisms that are being currently (and successfully) used to that
end actually imply the opposite choice ($\zeta=-1$). These include
at least the Bona-Mass\'o \cite{Bona92,Bona95}, Shibata-Nakamura
\cite{SN95}, Baumgarte-Shapiro \cite{BS99} and Anderson-York
\cite{Anderson99} formalisms. The reason may be that with the 
$\zeta=-1$ choice the transverse traceless part of the evolution system
looks like that of the wave equation, as pointed out in \cite{Cornell01},
which is known to admit a symmetric-hyperbolic form.

Symmetric hyperbolicity is a stronger requirement than just strong
hyperbolicity in order to ensure existence, unicity and stability of
the solutions arising from a given set of initial data (well posedness)
\cite{Kreiss89}. When the evolution system is just strongly hyperbolic,
but not symmetric hyperbolic, then well posedness is not ensured for
non-smooth data, like the ones arising in numerical experiments due
to finite machine accuracy.

We actually have some preliminary evidence that the choice $\zeta=+1$
is prone to numerical instabilities. To confirm this, one must proceed
to a systematic exploration of parameter space like the one presented in
the KST paper \cite{Cornell01} for the $\zeta=-1$ case. There is also
some theoretical evidence  \cite{Lee02} that the choice $\zeta=+1$,
when combined with conventional choices of the other parameters
introduced in \cite{Cornell01}, does not lead to a symmetric hyperbolic
system. This opens the way, from the theoretical point of view, to
using the symmetric hyperbolicity requirement, in the $\zeta=+1$ case,
to restrict the values of the other parameters, then reducing the volume
of parameter space.

We are currently working along these lines, both from the theoretical
and from the numerical point of view.\\\\

{\em Acknowledgements: We thank Manuel Tiglio for helpful and
stimulating discussions during his visit to Palma de Mallorca.
This work has been supported by the EU Programme
'Improving the Human Research Potential and the Socio-Economic
Knowledge Base' (Research Training Network Contract (HPRN-CT-2000-00137),
by the Spanish Ministerio de Ciencia y Tecnologia through the research
grant number BFM2001-0988 and by a grant from the Conselleria d'Innovacio
i Energia of the Govern de les Illes Balears.}

\bibliographystyle{prsty}

\end{document}